\documentclass[intlimits,twoside,a4paper]{article}

\usepackage{amsmath,amssymb,textcomp}
\usepackage{graphicx}

\usepackage[T2A]{fontenc}
\usepackage[cp1251]{inputenc}

\usepackage[eqsecnum]{cmpj2}

\usepackage{graphicx}  
\usepackage{bm}        
\usepackage{amssymb}   
\usepackage[percent]{overpic} 
\usepackage{wasysym}
\usepackage{hyperref}	

\issue{2017}{20}{1}{13701}
\doinumber{10.5488/CMP.20.13701}

\title[The large-$m$ limit, and kagome spin models]%
{The large-$m$ limit, and spin liquid correlations in kagome-like spin models}

\author[T. Yavors'kii]{T. Yavors'kii\thanks{E-mail: ab3785@coventry.ac.uk}}
\address{AMRC, Coventry University, CV1 5FB, United Kingdom}

\authorcopyright{T. Yavors'kii, 2017}
\date{Received January 26, 2017}

\DeclareMathOperator{\Tr}{Tr}

\begin{document}
\maketitle

\begin{abstract}
It is noted that the pair correlation matrix $\hat{\chi}$ of the nearest neighbor Ising model on
periodic three-dimensional ($d=3$) kagome-like lattices
of corner-sharing triangles can be calculated partially exactly.
Specifically, a macroscopic number $1/3 \, N+1$ out of $N$ eigenvalues of $\hat{\chi}$ are degenerate
at all temperatures $T$,
and correspond to an eigenspace $\mathbb{L}_{-}$ of $\hat{\chi}$, independent of $T$.
Degeneracy of the eigenvalues,
and $\mathbb{L}_{-}$ are an exact result for a complex $d=3$ statistical physical model.
It is further noted that the eigenvalue degeneracy
describing the same $\mathbb{L}_{-}$ is exact at all $T$ in an infinite spin dimensionality $m$ limit
of the isotropic $m$-vector approximation to the Ising models.
A peculiar match of the opposite $m=1$ and $m\rightarrow \infty$ limits
can be interpreted that the $m\rightarrow\infty$ considerations are exact for $m=1$.
It is not clear whether the match is coincidental.
It is then speculated that the exact eigenvalues degeneracy in $\mathbb{L}_{-}$ in the
opposite limits of $m$ can imply their quasi-degeneracy for intermediate $1 \leqslant m < \infty$.
For an anti-ferromagnetic nearest neighbor coupling,
that renders kagome-like models highly geometrically frustrated, these are spin states
largely from $\mathbb{L}_{-}$ that for $m\geqslant 2$ contribute to $\hat{\chi}$ at low $T$.
The $m\rightarrow\infty$ formulae can be thus quantitatively correct in description of $\hat{\chi}$
and clarifying the role of perturbations in kagome-like systems deep in the collective paramagnetic regime.
An exception may be an interval of $T$, where the order-by-disorder mechanisms
select sub-manifolds of $\mathbb{L}_{-}$.
\pacs 75.10.Hk, 05.50.+q, 75.25.+z, 75.40.Cx

\end{abstract}

\section{Introduction}
Exact insights into collective behaviors are rare even for
the simplest systems of many interacting constituents.
For instance, no exact solutions of short-range classic-spin models
on periodic lattices
are known above spatial dimension $d=2$
\cite{baxter2007exactly}.
An exception is critical behaviors above upper critical dimensions,
due to methods of the renormalization group
\cite{amit1984field}.
Properties of $d=3$ spin models are deduced from approximations.
A classic approximation is a limit
of the large spin-space dimension $m$,
and expansions about it \cite{brezin1993large}.

The approximation renders the models
trivially coupled and thus exactly solvable.
Yet these are non-trivial interactions and correlations between degrees of freedom that are behind the complexity of the $d=3$ physical world.
In this context,
highly geometrically frustrated (``frustrated'') condensed matter systems have recently attracted much attention.
At temperatures $T$ lower than the scale of the leading interactions $J$, they can form massively degenerate
highly correlated states, called spin liquids or collective paramagnets.
Exotic properties of these states are
believed behind the richness of quantum and classical phenomena observed in the frustrated systems
at low $T$ \cite{lacroix2011introduction}.

This paper presents observations regarding the characteristics of correlations in a family of
prototypical frustrated spin models on $d=2$ and $d=3$ kagome-like lattices.
First, both for $d=2$, and $d=3$ models, the correlations can be described exactly to a large extent.
Second, despite the models can form a collective paramagnetic phase,
the correlations in this phase can be well reproduced by the $m\rightarrow\infty$ approximation.
The paper suggests that, potentially with the exception of the $T$ interval where order-by-disorder mechanisms \cite{villain1980order}
are relevant,
the structure of the equilibrium pair correlation function
does not support the division into two distinct regimes of a paramagnet
and a collective paramagnet.
The kagome-like models thus may be ``transparent'' to the conventional paramagnetic treatment {\it deep below}
the mean-field critical temperature $\Theta_{\text c}$
that is usually interpreted as signaling the onset of a collective paramagnetic regime.

\section{Main result}
\begin{figure*}[!b]
\begin{overpic}[trim=230mm 10mm 180mm 70mm,clip,width=0.33\textwidth]{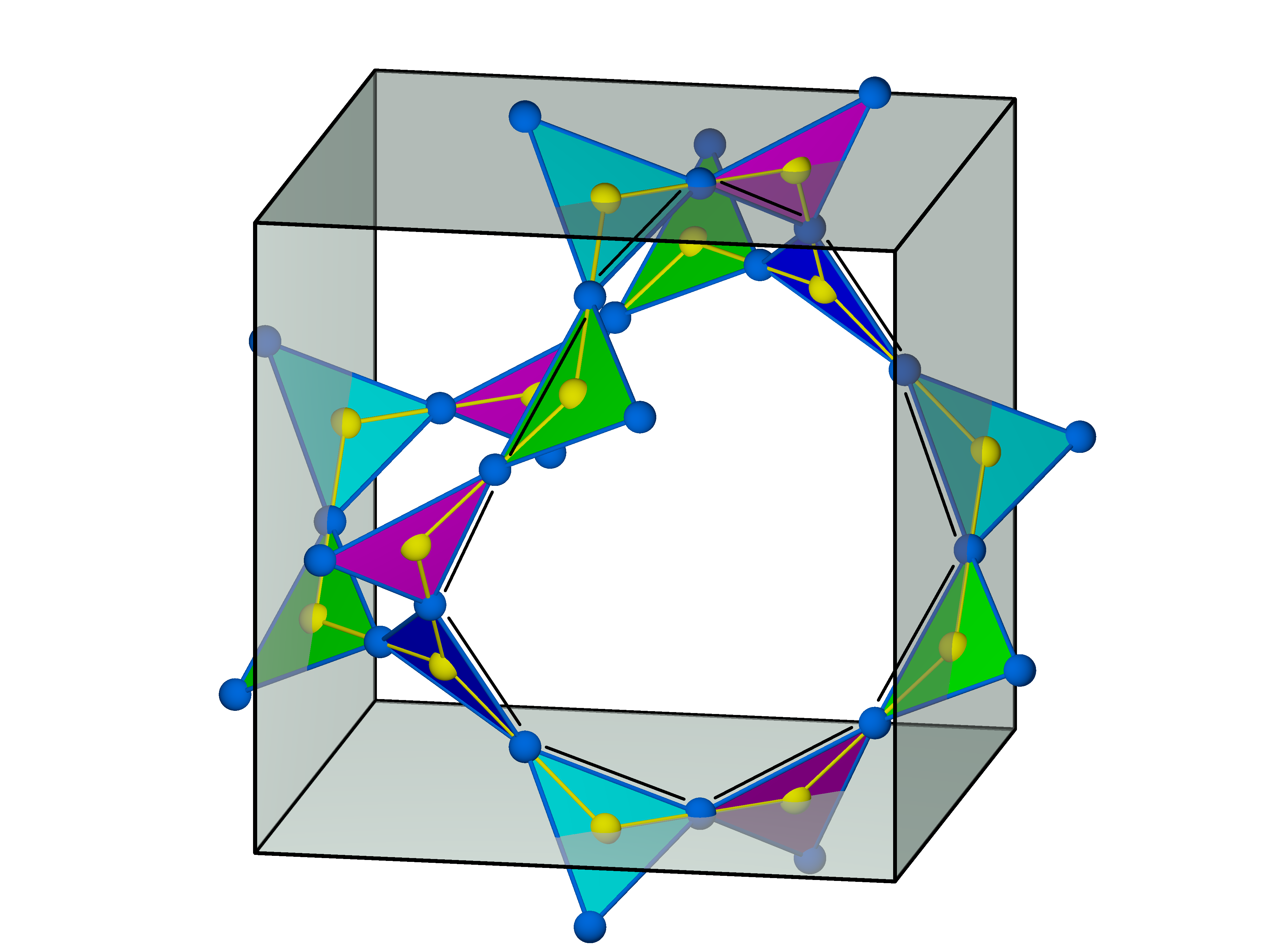}
\put(0,0) {\LARGE (a)}
\end{overpic}
\begin{overpic}[trim=25mm 95mm 25mm 25mm,clip,width=0.3\textwidth]{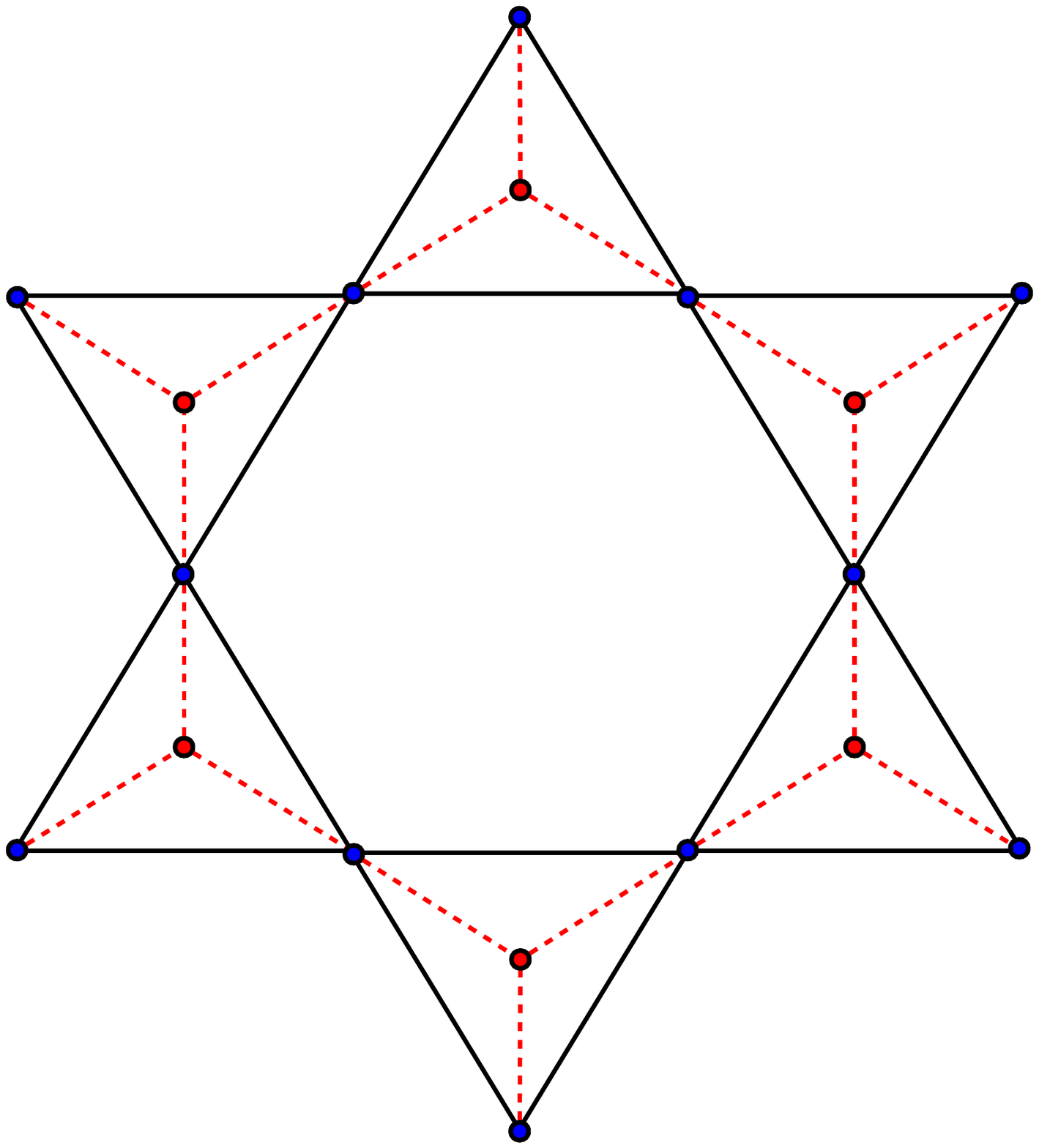}
\put(0,0) {\LARGE (b)}
\end{overpic}
\begin{overpic}[trim=-30mm 15mm 0mm 15mm,clip,angle=90,width=0.33\linewidth]{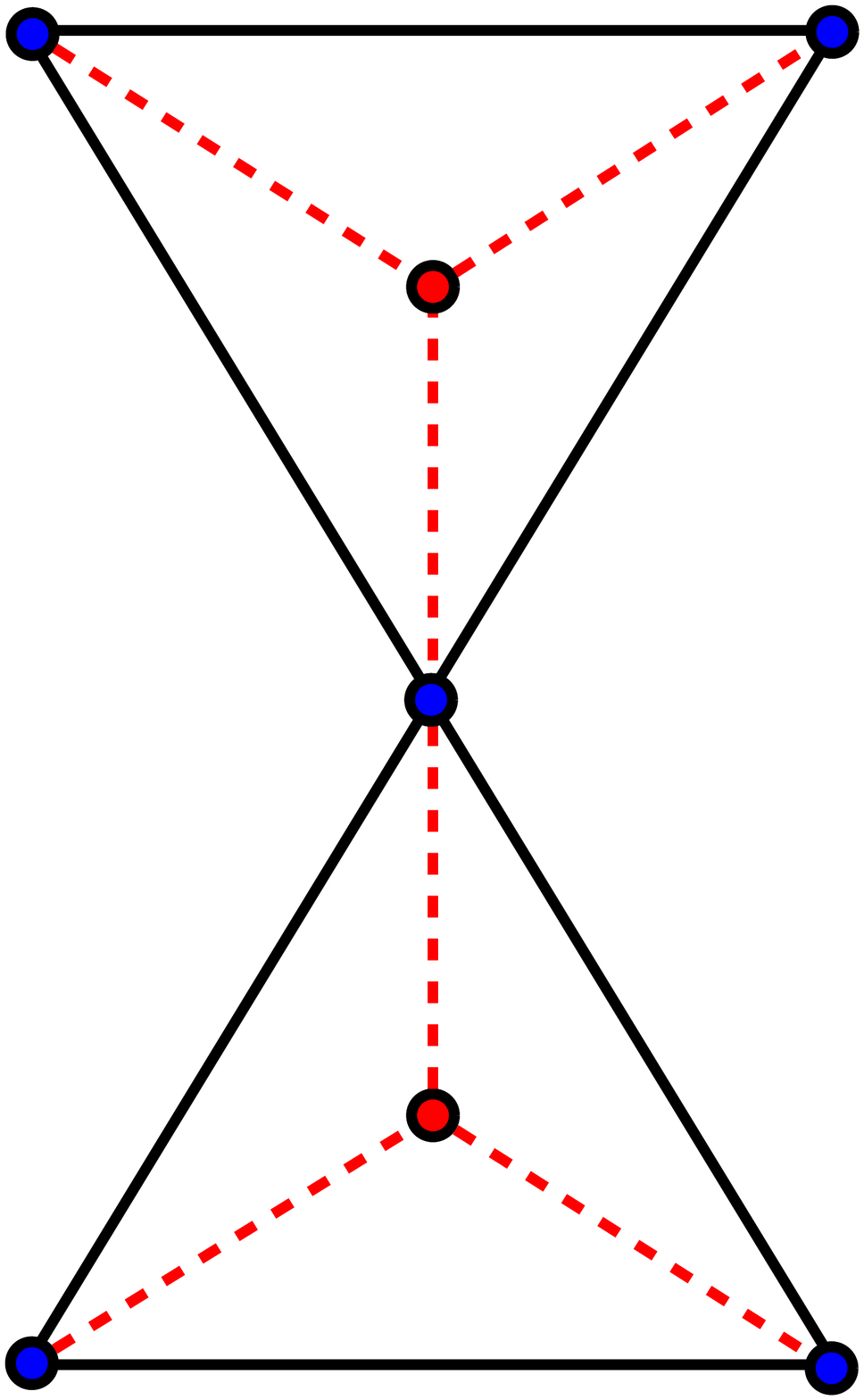}
\put(10,18) {\LARGE$\mu_1$}
\put(47,40) {\LARGE$\mu_2$}
\put(10,76) {\LARGE$\mu_3$}
\put(21,40) {\LARGE$\sigma$}
\put(74,40) {\LARGE$\sigma^{\prime}$}
\put(0,0) {\LARGE (c)}
\end{overpic}
\caption{\label{fig:lattices} (Color online)
(a) Kagome-like lattices
are periodic arrangements of points in the three-dimensional ($d=3$) space.
They are analogous to the kagome lattice in $d=2$ space (b).
(a, b) Each point is at the common corner of two equilateral triangles and
is in the same environment of other points.
Centers of triangles form the hexagonal
(b) and a hexagonal-like (a) lattices.
The kagome-like lattice (a) can be obtained by replicating a cubic unit cell,
similar to the simple cubic lattice;
unlike it, its unit cell is non-Bravais and consists of 12 points each.
(a) The super-imposed black piecewise line shows the shortest closed path on the kagome-like lattice that is allowed to
pass through one edge of each triangle only.
(c) Gibbs factor of three coupled Ising spins $\mu_1$, $\mu_2$, $\mu_3$ can be generated with the help of
an auxiliary spin $\sigma$, see equation~(\ref{eq:YDtrasformation}) in the main text.
Each site of a kagome-like lattice has two unique neighbors
$\sigma$ and $\sigma^{\prime}$ on a hexagonal-like lattice.
}
\end{figure*}

The paper is based on extending, connecting and interpreting two known observations {\it (1)} and {\it (2)} below.
Observation {\it (1)} is essentially due to \cite{huse1992classical,Rutenberg_1993};
observation {\it (2)} stems
from \cite{garanin1999classical,enjalran2004theory,isakov2004dipolar,yavors2006spin,hopkinson2007classical}.
Consider a lattice consisting of equivalent corner sharing triangles.
This can be a $d=2$ kagome lattice, or a $d=3$ kagome-like lattice,
shown in figure~\ref{fig:lattices}.
Other lattices, for which the argument of the paper holds,
can be seen e.g., in figure~2~(b), (c) of \cite{ishikawa2014kagome},
or in figure~1 of \cite{hopkinson2007classical}.
Kagome-like lattices describe magnetic materials.
For example, in gadolinium gallium garnet,
magnetic Gd$^{3+}$ ions occupy sites of two inter-penetrating species of the lattice of figure~\ref{fig:lattices}~(a),
which are separated by a distance larger than the n.n. distance in each species.
See, e.g., \cite{paddison2015hidden} for a list of positions of Gd$^{3+}$ in the cubic unit cell.

\begin{figure}[!t]
\begin{overpic}[trim=30mm 10mm 80mm 10mm,clip,angle=-90,width=0.49\textwidth]{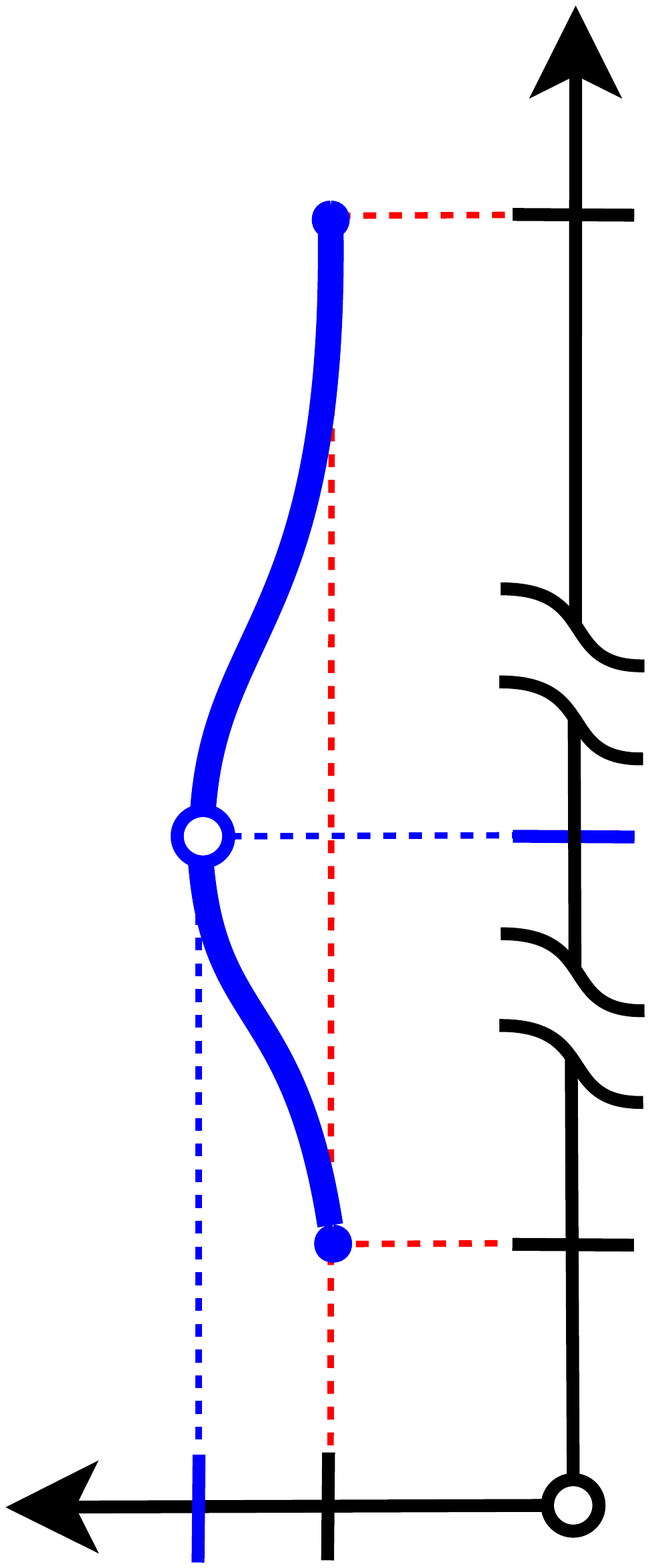}
\put(-3,0) {\LARGE (a)}
\put(87,10) {\LARGE$m$}
\put(26,8) {\LARGE$1$}
\put(71,8) {\LARGE$\infty$}
\put(14,31) {\LARGE$\Delta$}
\put(2,16) {\LARGE$0$}
\put(33,29) {\LARGE maximum}
\end{overpic}
\begin{overpic}[trim=25mm 5mm 80mm 5mm,clip,,angle=-90,width=0.49\textwidth]{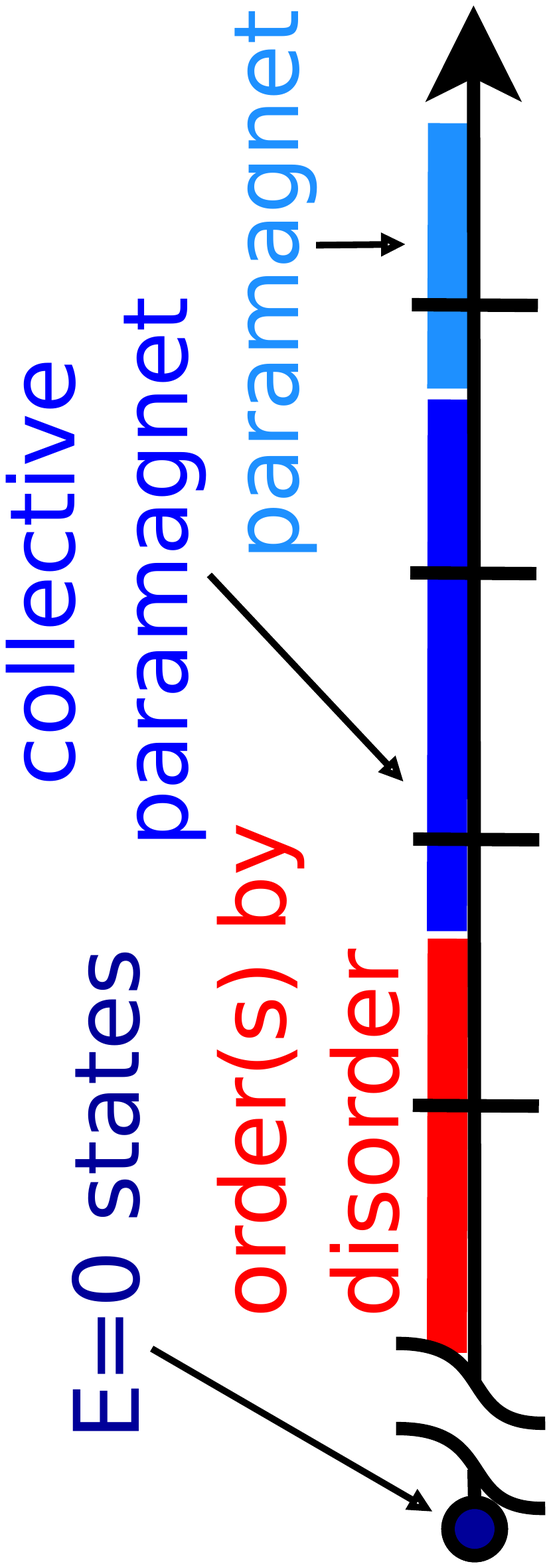}
\put(-1,0) {\LARGE (b)}
\put(82,0) {\LARGE$T/J$}
\put(74,0) {\LARGE$1$}
\put(57,0) {\LARGE$10^{-1}$}
\put(40,0) {\LARGE$10^{-2}$}
\put(23,0) {\LARGE$10^{-3}$}
\end{overpic}
\caption{\label{fig:diagrams} (Color online)
(a) Consider eigenvalues of the $N\times N$ correlation matrix $\hat{\chi}$
of a nearest neighbor antiferromagnetic $m$-vector
model on a $d$-dimensional kagome-like lattice in descending order
[see equation~(\ref{eq:chi_k}) and the following text]. The difference $\Delta$ between the eigenvalues number $1$ and
$1/3 \, N +1$ is exactly zero in two limits, the Ising limit $m=1$ and the spherical limit $m\rightarrow\infty$.
High-$T$ expansion \cite{harris1992possible} suggests non-zero $\Delta$ for other values of $m$.
The magnitude of $\Delta$ can be thought of as a measure of deviations of correlations
from their $m\rightarrow\infty$ form.
Small deviations can explain the heuristics of applicability of the
$m\rightarrow\infty$ limit, and the variational mean-field theory, to study correlations
deep in the collective paramagnetic regimes
\cite{enjalran2004theory,yavors2006spin,isakov2004dipolar,hopkinson2007classical}
of the finite-$m$ models.
(b) The behavior of n.n. classical $m$-vector models on regular frustrated lattices, such as the $d=2$ kagome lattice,
can be divided into four regimes. As temperature $T$ is lowered, a model
can firstly cross-over from a high-$T$ paramagnetic phase to
a correlated collective paramagnetic phase at about the mean-field critical temperature $\Theta_{\text c}$.
Then, the model can undergo (a sequence of) cross-overs or phase transitions due to
order-by-disorder mechanisms \cite{villain1980order} that are activated at temperatures $\Theta_{\rm obd}\ll\Theta_{\text c}$.
Strictly at $T\equiv0$, there can be a discontinuity, that separates the
$T\rightarrow 0$ phase from the microcanonical ground states phase.
The paper suggests that the juxtaposition of $T\equiv0$, collective paramagnetic and paramagnetic phases
as separated entities may be not supported by the structure of the equilibrium correlation function.
The exception might be the regime $0<T\lesssim\Theta_{\rm obd}$ where the order-by-disorder is important.
The diagram is qualitative and does not show exact energy scales.
}
\end{figure}

Define an isotropic $m$-vector model on the lattice of figure~\ref{fig:lattices}~(a) by the Hamiltonian:
\begin{equation}
\mathcal{H}=\sum_{i,j} J(i,j) \, ({\bm\mu}_i \cdot {\bm\mu}_j)\,,
\label{eq:H_k}
\end{equation}
where $i,j$ span $N$ sites of the lattice,
each site $i$ carries an $m$-dimensional isotropic $O(m)$ vector spin ${\bm\mu}_i$ of length $\sqrt{m}$,
the spins are coupled via a dot product,
and entries of the symmetric interaction matrix $J(i,j)$ are $0$, except for the nearest neighbor (n.n.) sites,
when they are half the n.n. coupling $J=1$.
Let $\langle \cdots \rangle \, (\beta)$ denote a Gibbs ensemble average defined by (\ref{eq:H_k}).
For instance, a spin-spin correlation matrix $\hat{\chi}^{\mu\nu}$ reads:
\begin{equation}
\chi^{\mu\nu}(i,j) = \langle \mu_i^{\mu} \mu_j^{\nu} \rangle =
\frac{ \Tr_{\mathbb{\mu}} \mu_i^{\mu} \mu_j^{\nu} \exp(-\beta\mathcal{H}) }{ \Tr_{\mu} \exp(-\beta\mathcal{H}) }\,,
\label{eq:chi_k}
\end{equation}
where
$\mu$, $\nu$ enumerate the components of spins,
$\beta=1/T$ is the inverse temperature, and $\Tr_{\mu}$ means integration over all degrees of freedom.
We assume $\hat{\chi}^{\mu\nu}=\hat{\chi}\times\delta^{\mu\nu}$ with
$\delta^{\mu\nu}$ being Kronecker delta.
The assumption would hold for phases that preserve the global spin rotational symmetry of the Hamiltonian,
for example in the paramagnetic, collective paramagnetic and $E=0$ phases [cf. figure~\ref{fig:diagrams}~(b)].
In this way, any $\hat{\chi}^{\mu\nu}$ is fully characterized by the matrix $\hat{\chi}$,
whose dimensions are $N\times N$ independently of $m$. Below, we are interested in the properties of $\hat{\chi}$ as a function of $m$.

For any finite model (\ref{eq:H_k}) of $L^3$ cubic unit cells with periodic boundary conditions,
the following two statements about $\hat{\chi}$ are correct.

{\it (1)} For the Ising case $m=1$, $\mu_i= \pm 1$,
the macroscopic number $1/3 \, N+1$ out of $N$ eigenvalues of $\hat{\chi}$ coincide (are degenerate)
at any $\beta$. At $\beta>0$, they are the largest eigenvalues in the spectrum of $\hat{\chi}$.
Here, $N=12 L^3$ is the number of spins in the model.

The eigenspace $\mathbb{L}_{-}$ of the degenerate eigenvalues of $\hat{\chi}$ is solely determined
by the interaction matrix $\hat {J}$ of (\ref{eq:H_k}) and is independent of $\beta$.
Specifically, $\mathbb{L}_{-}$ coincides with the eigenspace of $\hat {J}$ of dimension $1/3 N+1$ corresponding to its degenerate
minimal eigenvalue $-1$. Informally:
\begin{equation}
\hat {J} \mathbb{L}_{-} = - \mathbb{L}_{-}\,.
\label{eq:L_}
\end{equation}

{\it (2)} For the case $m\rightarrow\infty$ at $\beta>0$,
the macroscopic number $1/3 \, N+1$ of the largest eigenvalues of $\hat{\chi}$ are again
degenerate and describe the same $\mathbb{L}_{-}$ (\ref{eq:L_}).

\section{Derivation}
{\it (1)}
Consider the Ising version $m=1$ of (\ref{eq:H_k}).
A star-triangle ($Y-\Delta$) transformation, said to be due to Onsager,
relates exactly the zero-field partition function
of n.n. Ising models on $d=2$ kagome, hexagonal and triangular lattices \cite{syozi1951statistics}.
A perhaps less known its application is a relationship between the $n$-spin correlation functions
of the three models \cite{barry1988exact}.
In particular, \cite{huse1992classical,Rutenberg_1993} showed that
the largest eigenvalues of the correlation matrix of the $d=2$ kagome Ising model are degenerate at all $T$.
The argument uses a local lattice topology and works for Ising models on lattices at any $d$,
as soon as they consist of corner sharing triangles.
This paper adopts the argument \cite{huse1992classical,Rutenberg_1993,barry1988exact}
to three-dimensional lattices, such as the kagome-like lattice of figure~\ref{fig:lattices}~(a).

The $Y-\Delta$ transformation recasts the Boltzmann weight factor of any three
coupled Ising variables $\mu_1$, $\mu_2$, $\mu_3$, figure~\ref{fig:lattices}~(c),
via a partial summation $\Tr_{\sigma}$ over a new Ising variable $\sigma$:
\begin{equation}
\Tr_{\sigma} \re^{-\beta_d \sigma(\mu_1+\mu_2+\mu_3)} \,=\, A \re^{-\beta \,(\mu_1 \mu_2+\mu_2 \mu_3 + \mu_3 \mu_1)} \,.
\label{eq:YDtrasformation}
\end{equation}
Here, $\beta_d$ and $A$ are known functions of $\beta$.
Introduction of $\sigma$ decouples $\mu_1$, $\mu_2$, $\mu_3$.
We use new variables $\{\sigma\}$ to decouple all triangle-coupled kagome spins $\{\mu\}$,
and then sum $\{\mu\}$ out
(applying the ``decoration-iteration'' transformation \cite{barry1988exact}).
Graphically, the remaining variables $\{\sigma\}$ can be thought of as
forming a n.n. Ising model on a lattice of the centers of the original corner-sharing triangles:
the hexagonal lattice at $d=2$ and a hexagonal-like lattice at $d=3$, see figure~\ref{fig:lattices}~(a), (b),
with the spin number $N_{\,\text h}=2/3 N$.
Analytically, the partition functions $Z_{\,\text h}$ and $Z$ of {\it any} pair of the hexagonal-like and kagome-like
Ising models become related by the same, exact formula of \cite{syozi1951statistics}.

To recast the kagome-like model spin correlations in terms of
the hexagonal-like model spin correlations,
we again decouple $\{\mu\}$
in the Boltzmann factors by using $\{\sigma\}$.
However, when summing $\{\mu\}$ out, we include a product of the chosen $\mu_j \mu_j$ in (\ref{eq:chi_k}).
Same procedure works for multi-spin correlations.
Note that every spin $\mu_i$ of the kagome-like lattice neighbors a unique pair of spins,
say $\sigma_{i}$, $\sigma_{i}'$, that are nearest neighbors on the corresponding hexagonal-like lattice,
cf. figure~\ref{fig:lattices}~(c).
Denoting $\bar{\mu}_i=\sigma_i + \sigma_i^{\prime}$, we obtain~\cite{huse1992classical,Rutenberg_1993}:
\begin{equation}
\langle \mu_i \mu_j \rangle \, (\beta) =
\delta_{ij} \, \left(1+M^2 \langle \bar{\mu}_i^2 \rangle_{\text h}\right) \,- \,
M^2 \, \langle \bar{\mu}_i \bar{\mu}_j \rangle_{\text h} \,.
\label{eq:corr-correspondance}
\end{equation}
Here,
$\langle \cdots \rangle_{\text h} \, (\beta_{\,\text h})$ means thermal average in
the hexagonal-like model with n.n. coupling $1$ at the inverse temperature $\beta_{\,\text h}$,
and
$
M^2(\beta_{\,\text h})=1/4 \, (\re^{4\beta_{\,\text h}}-1)>0\,.
$

Consider a sign-alternating linear combination $\tau$ of kagome spins lying on a closed path.
For the kagome-like lattice of figure~\ref{fig:lattices}~(a), the mode $\tau$ can
be formed of ten spins
$\mu_1,\ldots,\mu_{10}$
of the loop in figure~\ref{fig:lattices}~(a):
$\tau=c_{1} \mu_{1} + \cdots + c_{10} \mu_{10}$,
where $c_1=1$, $c_2=-1$, $\ldots\,$, $c_{10}=-1$.
Note that (\ref{eq:corr-correspondance}) is a difference of an identity matrix times a constant,
and a positive semi-definite matrix.
In (\ref{eq:corr-correspondance}), the special choice of $\tau$ zeros the second term contribution
to the mode susceptibility $\langle \tau^2 \rangle$, thus
maximizing it.
This makes any $\vec{c}=(\cdots,c_k,\cdots)$,
whose entries are non-zero only if they coincide with the sign-alternating coefficients
of a loop, the eigenvector corresponding to the largest eigenvalue of $\hat{\chi}$.

Observe that (\ref{eq:H_k}) can be written as a sum of squares minus a constant.
It is clear that every $\vec{c}$ zeros the squares,
and thus is the eigenvector of $\hat{J}$ corresponding to the smallest eigenvalue $-1$ of $\hat{J}$.
The linear span of all $\vec{c}$ forms $\mathbb{L}_{-}$,
whose dimension is $N$ minus the dimension of the triangles constraints in the sum of squares, which
is $N_{\,\text h}-1$. We have: $\dim \mathbb{L}_{-} = 1/3\,N +1$.

{\it (2)}
Examine (\ref{eq:H_k}) at arbitrary $m$.
\cite{stanley1968spherical,angelescu1979spherical} showed
that an $m$-vector lattice model is exactly solvable in the limit $m \rightarrow \infty$,
where it coincides with the spherical model of Berlin and Kac \cite{berlin1952spherical}.
In particular, the spherical limit of the correlation matrix (\ref{eq:chi_k})
of model (\ref{eq:H_k}) reads \cite{angelescu1979spherical}:
\begin{equation}
\hat{\chi}_{\ocircle} = \frac{1}{2}\frac{1}{r_0+\beta \hat{J}}\,,
\label{eq:chi_k_O}
\end{equation}
where parameter $r_0$ is fixed by normalization
\begin{equation}
\Tr  \frac{1}{2} \frac{1}{r_0+\beta \hat{J}} = N\,,
\label{eq:O_normalization}
\end{equation}
and $\hat{J}$ is the interaction matrix in (\ref{eq:H_k}).
The variational mean-field theory \cite{chaikin2000principles}, see e.g., \cite{enjalran2004theory}
for its application in the context
of frustrated magnetism,
gives the dependence of $\hat{\chi}$ on $\hat{J}$ in the same form
of a $[0/1]$ Pad\'e approximant.
Since $\hat{\chi}_{\ocircle}$ is an (analytic) function of $\hat{J}$, $\hat{\chi}_{\ocircle}$
has a set of degenerate eigenvalues corresponding to $\mathbb{L}_{-}$.
Since $\hat{\chi}_{\ocircle}$ is a monotonously decreasing function of $\hat{J}$, $\beta>0$,
the smallest degenerate eigenvalues in the spectrum of $\hat{J}$ are the largest
in the spectrum of $\hat{\chi}_{\ocircle}$.

\section{Interpretation}
It was previously observed, for instance in \cite{isakov2004dipolar,hopkinson2007classical},
that the $m\rightarrow\infty$ formulae provide an excellent fit to the collective paramagnetic correlations of finite-$m$
n.n. $m$-vector models on the $d=3$ pyrochlore and a kagome-like frustrated lattice.
The variational mean-field theory
was observed in e.g., \cite{enjalran2004theory,yavors2006spin,henelius2016refrustration} to quantitatively correctly describe
the role of perturbations in lifting the degeneracy in the collective paramagnetic regimes.
This paper may shed light as to
why these theories are well applicable into collective paramagnetic regimes
for the case of kagome-like lattices.\footnote{
Pyrochlore lattice case will be presented elsewhere.
}

We showed above that the upper $1/3\,N+1$ eigenvalues
of the correlation matrix $\hat{\chi}$ (\ref{eq:chi_k})
of model (\ref{eq:H_k}) are exactly degenerate for $m=1$ and $m\rightarrow\infty$ at all $T$,
and the corresponding eigenspace $\mathbb{L}_{-}$ is independent of $T$.
We can quite naturally conjecture that
for the intermediate values $1 \leqslant m < \infty$, the eigenvalues of $\mathbb{L}_{-}$ become at all $T$ only weakly dispersed.
The upper eigenvalues can remain quasi-degenerate also for $m<1$, for instance in the polymer limit $m\rightarrow 0$.
Observe that at $T\equiv 0$, i.e., strictly in the phase of the microcanonical ground states,
and at $m \geqslant 2$, these are the spin states belonging to $\mathbb{L}_{-}$ only
that contribute to $\hat{\chi}$.
Thus, at $T\equiv 0$, $\hat{\chi}$ is determined by its approximate form of the $m\rightarrow\infty$ projector on
the linear space~$\mathbb{L}_{-}$.

We can consider the (relative) dispersion $\Delta$ of the
quasi-degenerate eigenvalues of $\hat{\chi}$
as a measure of deviations
of correlations from the $m\rightarrow\infty$ projector form.
The dependence of $\Delta$ on $m$ can have a shape of figure~\ref{fig:diagrams}~(a).
The location $m_0$ of a maximum
might depend on $d$ and on the choice of the kagome-like lattice, but might not exceed $3$.
For instance, no order-by-disorder phenomenology was observed for larger $m$ on the $d=2$ kagome lattice
\cite{huse1992classical,Rutenberg_1993}, pointing that such a model is in the $m\rightarrow\infty$ regime,
where no order-by-disorder is observed either.

We can next speculate that the $m\rightarrow\infty$ projector form $\Delta\approx 0$ of $\hat{\chi}$ is valid
in the collective paramagnetic phase at a finite $T$, which by definition mainly consists of
the states from the extensively degenerate manifold $\mathbb{L}_{-}$.
The projector form would hold several orders of magnitude in $T$
below $\Theta_{\text c}$, the mean-field critical temperature,
but potentially above $\Theta_{\rm obd}$, the order-by-disorder temperatures,
where thermal fluctuations can select subset(s) of $\mathbb{L}_{-}$ with the greatest number of low-energy excitations
\cite{shender1982antiferromagnetic,henley1987ordering},
cf. figure~\ref{fig:diagrams}~(b).
As $\mathbb{L}_{-}$ is known exactly,
the interesting question about the structure of correlations
at low $T$ may be not the projector form of $\hat{\chi}$ per se,
but the nature of the (small) deviations from it.
Above about $\Theta_{\text c}$, the $m\rightarrow\infty$ form (\ref{eq:chi_k_O}) can be expected to apply naturally.
Therefore, the correlations in model (\ref{eq:H_k}) can be well reproducible by
their $m\rightarrow\infty$ form for $m\geqslant2$ and at all $T$,
with the potential exception of the phases dictated by the order-by-disorder.

Consider any other Hamiltonian $\mathcal{H}^{\prime}$ on a kagome-like lattice, which preserves the
symmetry of the lattice.
Let, for $m\geqslant2$, $\mathcal{H}^{\prime}$
admit the same microcanonical $T\equiv 0$ degenerate ground states as the original ${\cal H}$ (\ref{eq:H_k}). For instance,
${\cal H}^{\prime}$ can be obtained from (\ref{eq:H_k}) by using another interaction matrix $\hat{J}^{\prime}$,
for which (\ref{eq:L_}) is true,
but which is not necessarily the nearest neighbor.
The coincidence of ground states for distinct ${\cal H}$  and  ${\cal H}^{\prime}$
was a dubbed projective equivalence in \cite{isakov2005spin} in the context of spin models
on the pyrochlore lattice.
As $T\equiv 0$ states of ${\cal H}^{\prime}$ and ${\cal H}$ coincide,
the upper eigenvalues of the correlation matrix are again quasi-degenerate for all $m\geqslant 2$.
The quasi-degeneracy, and the $m\rightarrow\infty$ form of correlations for ${\cal H}^{\prime}$
should be again correct for $m\geqslant2$ and all $T$, potentially excluding the window $0<T\lesssim\Theta_{\rm obd}$.

Consider another Hamiltonian ${\cal H}^{\prime\prime}$
different from ${\cal H}^{\prime}$ by small perturbations such that (\ref{eq:L_}) is not valid.
The perturbations can force the model ${\cal H}^{\prime\prime}$ to undergo a phase transition
at $T>\Theta_{\rm obd}$, the regime where equation~(\ref{eq:chi_k_O}) is applicable.
We can thus use equation~(\ref{eq:chi_k_O}) to study, for instance, the selection of the ordering wave vectors
dictated by perturbations. In essence, frustration may be unimportant for applicability of
the $m\rightarrow\infty$ approximation, while the variational mean-field theory may be used
for the study of the role of perturbations deep below the mean-field critical
temperature $\Theta_{\text c}$ \cite{enjalran2004theory,yavors2006spin,isakov2004dipolar,hopkinson2007classical}.
If we regard the applicability of these approaches as defining the nature of correlations,
there may be no difference between a collective and regular paramagnet.
Correspondingly, the $m\rightarrow\infty$, and the related variational mean-field approaches
may claim back their status as simple, powerful and standard tools for the study of
perturbations at low $T$ in kagome-like and other frustrated systems,
as was heuristically observed for instance in~\cite{enjalran2004theory,yavors2006spin}.

\newpage
\section*{Acknowledgements}
The author expresses his gratitude to Michel Gingras, multiple discussions with whom
inspired this project; thanks Andrew Rutenberg for a copy of his PhD thesis;
acknowledges useful discussions with Peter Holdsworth, Wolfhard Janke, Martin Weigel,
Nikolay Izmailian, Yurij Holovatch, Volodymyr Tkachuk, Tom Fennell.

\newpage
\ukrainianpart

\title[Велика вимірність спіна, та каґомеподібні спінові моделі]%
{Границя великої вимірності спіна та кореляції спінової рідини у каґомеподібних спінових моделях}
\author{Т. Яворський}
\address{Дослідницький центр прикладної математики, університет Ковентрі, CV1 5FB, Великобританія}

\makeukrtitle

\begin{abstract}
Зауважено, що парнокореляційну матрицю $\hat{\chi}$ моделі Ізинґа найближчих сусідів
на періодичних тривимірних ($d=3$) каґомеподібних гратках можна обчислити частково точно.
Зокрема, $1/3 \, N+1$ із $N$ власних значень $\hat{\chi}$ вироджені
при усіх температурах $T$ та відповідають власному лінійному простору $\mathbb{L}_{-}$ матриці $\hat{\chi}$,
незалежному від $T$. Виродження власних значень
та $\mathbb{L}_{-}$ --- приклад точного результату для складної $d=3$ моделі статистичної фізики.
Зауважено далі, що виродження власних значень, які описують той самий $\mathbb{L}_{-}$, ---
точне при усіх $T$ у границі безмежної вимірності спіна $m$,
яку можна розглядати як наближення ізотропної $m$-векторної моделі до моделі Ізинґа.
Своєрідне співпадіння протилежних $m=1$ та $m\rightarrow \infty$ границь
можна проінтерпретувати у спосіб, що міркування для $m\rightarrow\infty$ залишаються точними при $m=1$.
Незрозуміло, чи співпадіння випадкове.
Накінець зроблено припущення, що точне виродження власних значень у $\mathbb{L}_{-}$
у протилежних границях $m=1$ та $m\rightarrow \infty$ може означати їх
квазівиродження при $1 \leqslant m < \infty$.
Для антиферомагнітної константи зв'язку між найближчими сусідами,
при якій каґомеподібні моделі стають сильно геометрично фрустрованими, саме стани
із $\mathbb{L}_{-}$ роблять переважний внесок у $\hat{\chi}$ при низькій $T$ для $m\geqslant 2$.
Це означає, що рівняння у границі $m\rightarrow\infty$ можуть бути чисельно правильні для опису $\hat{\chi}$
та уточнення ролі збурень у каґомеподібних системах глибоко у режимі колективного парамаґнетика.
Винятком може бути інтервал $T$, де механізми лад-безлад вибирають підпростори $\mathbb{L}_{-}$.

\keywords гратка каґоме, фрустрація, спінові кореляції, точний результат
\end{abstract}

\end{document}